\documentclass[twocolumn]{aastex63}

\received{}
\revised{}
\accepted{}

\submitjournal{ApJ}

\shortauthors{L\'{o}pez {\it et al.}}
\usepackage{amsmath}
\usepackage{color}
\usepackage{natbib}
\usepackage{bm}
\begin{document}
%
\title{Alternative  high plasma beta regimes of electron heat-flux instabilities in the  solar wind}
%
\correspondingauthor{R.~A. L\'{o}pez}
\email{rlopez186@gmail.com}

\author[0000-0003-3223-1498]{R.~A. L\'{o}pez}
\affiliation{Departamento de F\'{i}sica, Universidad de Santiago de Chile, Casilla 307, Santiago, Chile}

\author[0000-0002-8508-5466]{M. Lazar}
\affiliation{Centre for mathematical Plasma Astrophysics,
  KU Leuven, Celestijnenlaan 200B, B-3001 Leuven, Belgium}
\affiliation{Institut f\"{u}r Theoretische Physik, Lehrstuhl IV:
  Weltraum- und Astrophysik, Ruhr-Universität Bochum, D-44780 Bochum,
  Germany}

\author[0000-0003-0465-598X]{S.~M. Shaaban}
\affiliation{Centre for mathematical Plasma Astrophysics, KU Leuven,
  Celestijnenlaan 200B, B-3001 Leuven, Belgium}
\affiliation{Theoretical Physics Research Group, Physics Department,
  Faculty of Science, Mansoura University, 35516, Mansoura, Egypt}

\author[0000-0002-1743-0651]{S. Poedts}
\affiliation{Centre for mathematical Plasma Astrophysics, KU Leuven,
  Celestijnenlaan 200B, B-3001 Leuven, Belgium}
\affiliation{Institute of Physics, University of Maria Curie-Skłodowska, PL-20-031 Lublin, Poland}

\author[0000-0002-9161-0888]{P.~S. Moya}
\affiliation{Centre for mathematical Plasma Astrophysics, KU Leuven,
  Celestijnenlaan 200B, B-3001 Leuven, Belgium}
\affiliation{Departamento de F\'{i}sica, Facultad de Ciencias,
  Universidad de Chile, Santiago, Chile}

\begin{abstract}
The heat transport in the solar wind is dominated by the 
suprathermal electron populations, i.e., a tenuous halo and a field-aligned 
beam/strahl, with high energies and antisunward drifts along the magnetic field. 
Their evolution may offer plausible explanations for the rapid decrease of the 
heat flux with the solar wind expansion, typically invoked being the 
self-generated instabilities, or the so-called heat flux instabilities (HFIs). 
The present paper provides a unified description of the full spectrum of HFIs,
as prescribed by the linear kinetic theory for high beta conditions ($\beta_e \gg 0.1$) 
and different relative drifts ($U$) of the suprathermals.
HFIs of different nature are distinguished, i.e., electromagnetic, electrostatic or
hybrid, propagating parallel or obliquely to the magnetic field, etc., as well 
as their regimes of interplay (co-existence) or dominance. 
These alternative regimes of HFIs complement each other and may be characteristic 
to different relative drifts of suprathermal electrons and various conditions in the 
solar wind, e.g., in the slow or fast winds, streaming interaction regions 
and interplanetary shocks. Moreover, these results strongly suggest that heat 
flux regulation may not involve only one but several HFIs, concomitantly or 
successively in time. Conditions for a single, well defined instability with major 
effects on the suprathermal electrons and, implicitly, the heat flux, seem to be
very limited. Whistler HFIs are more likely to occur but only for minor drifts
(as also reported by recent observations), which may explain a modest implication 
in their regulation, shown already in quasilinear studies and numerical simulations.
%
%
\end{abstract}


\section{An introductory motivation}

The solar wind heat flux is mainly attributed to the energetic suprathermal electrons, a 
diffuse halo present at all pitch-angles, and an electron beam, or strahl, directed along
the interplanetary magnetic field away from the sun. Suprathermals may not exceed $10 \%$
of the total density, but have high energies (much higher than thermal or core electrons) 
and significant antisunward drifts \citep{Pilipp-etal-1987, Wilson-etal-2019b}. The strahl is in general responsible for a major velocity shift between the core and suprathermal electrons 
\citep{Rosenbauer-etal-1977, Pilipp-etal-1987, Wilson-etal-2019b}, but recent studies 
also reveal a relative drift of the halo \citep{Wilson-etal-2019b} to be taken into 
account in certain circumstances; for instance, in the low-speed winds the strahl can be 
almost absent \citep{Gurgiolo-Goldstein-2017} and the heat is transported by the halo 
electrons \citep{Pilipp-etal-1987, Pagel-etal-2005, Bale-etal-2013}. However, if the 
strahl is observed most of the solar wind heat flux is carried by the strahl electrons 
\citep{Pilipp-etal-1987, Pagel-etal-2005, Graham-etal-2017, Lazar-etal-2020}.

The modifications of suprathermal electrons with the solar wind expansion can be directly 
linked to the variations of heat flux, and are expected to explain the observed dropouts 
and an accelerated decrease of the heat flux, more rapid than predicted by an adiabatic
decrease of the main plasma parameters. Indeed, the observations reveal an important 
erosion of the strahl, which decline in relative density and drift, and broaden their 
pitch-angle distribution with increasing heliocentric distance \citep{Maksimovic-etal-2005, Pagel2007, Anderson-etal-2012, Graham-etal-2017, Bercic2019}. The effect of binary collisions on suprathermals is insignificant, but these evolutions may be explained by the so-called heat flux instabilities (HFIs), self-generated by the relative drifts and beaming velocity of suprathermal electrons \citep{Gary1977, Gary-etal-1999a, Pavan-etal-2013, Shaaban2018MN, Shaaban-etal-2019, Vasko2019, Verscharen2019b}. The resulting wave fluctuations can induce a diffusion of suprathermals in velocity space, contributing to their relaxation, as already shown in numerical simulations 
\citep{Dum-Nishikawa-1994, Gary-Saito-2007, Kuzichev2019, Lopez-etal-2019}.

In this paper we provide a comparative analysis of the full spectrum of HFIs prescribed 
by the linear kinetic theory for high plasma beta conditions ($\beta \gg 0.1$) and 
different relative drifts ($U$) of the suprathermal populations. Such a unified 
analysis offers new and multiple perspectives for the implication of HFIs in the 
evolution of suprathermals and, implicitly, of the solar wind heat flux. Current way 
of thinking that a single instability can be identified as the principal mechanism 
of regulation of the heat flux in the solar wind may need major upgrades, to include 
the interplay and/or succession of two or more instabilities. 

In section~\ref{sec:theory} we introduce the kinetic formalism often adopted in 
studies of plasma wave dispersion and stability, in our case a typical plasma with 
two asymmetric counter-drifting populations of electrons. A short description is 
also provided for the numerical solver allowing us to determine the full spectrum 
of the unstable solutions, covering all ranges of frequencies, wave-numbers and 
angles of propagation. Models assumed for the zero-th order velocity distribution 
are drifting-Maxwellian, which enable a standard and simple parameterization of 
the solar wind electron-proton plasma populations. We are aware of the existence 
of other more realistic representations, like Kappa models \citep{Shaaban2018MN} 
for the halo, or more asymmetric combinations of drifting Maxwellians for a more
skewed strahl \citep{Horaites2018}, which would only complicate our analysis but 
ultimately would lead to similar results and conclusions. Adopting drifting 
Maxwellian keeps at this stage the analysis simple and enables straightforward 
interpretations of the HFIs, their nature, interplay and dominance. 
Moreover, such a dual model can reproduce the slow wind core-halo distribution, 
in the absence of strahl, but may also be relevant for the fast wind 
core-strahl configuration if the less drifting halo is assimilated to 
the core population\footnote{High beta ($\beta_e > 0.5$) instabilities may not be
significantly altered by the inclusion of halo in this case \citep{Horaites2018}}. 
The results are presented and discussed in detail in section~\ref{sec:results}, 
considering each alternative unstable regime in part. These regimes have a wide relevance, covering lower drifts and higher thermal spreads reproducing better the halo electrons, or higher drifts and lower thermal spreads specific to the strahl population, and, nevertheless, a series of intermediary states which may be associated with the relaxation of strahl and the formation or/and enhancement of halo \citep{Hammond1996, Anderson-etal-2012, Graham-etal-2017}.
The last section summarizes our results and formulates a series of conclusions, which 
should help in understanding the observations and make realistic interpretations of HFIs and their implications.

\section{Dispersion and stability} \label{sec:theory}

\begin{figure*}[t!]
  \begin{center}
    \includegraphics[width=0.95\textwidth]{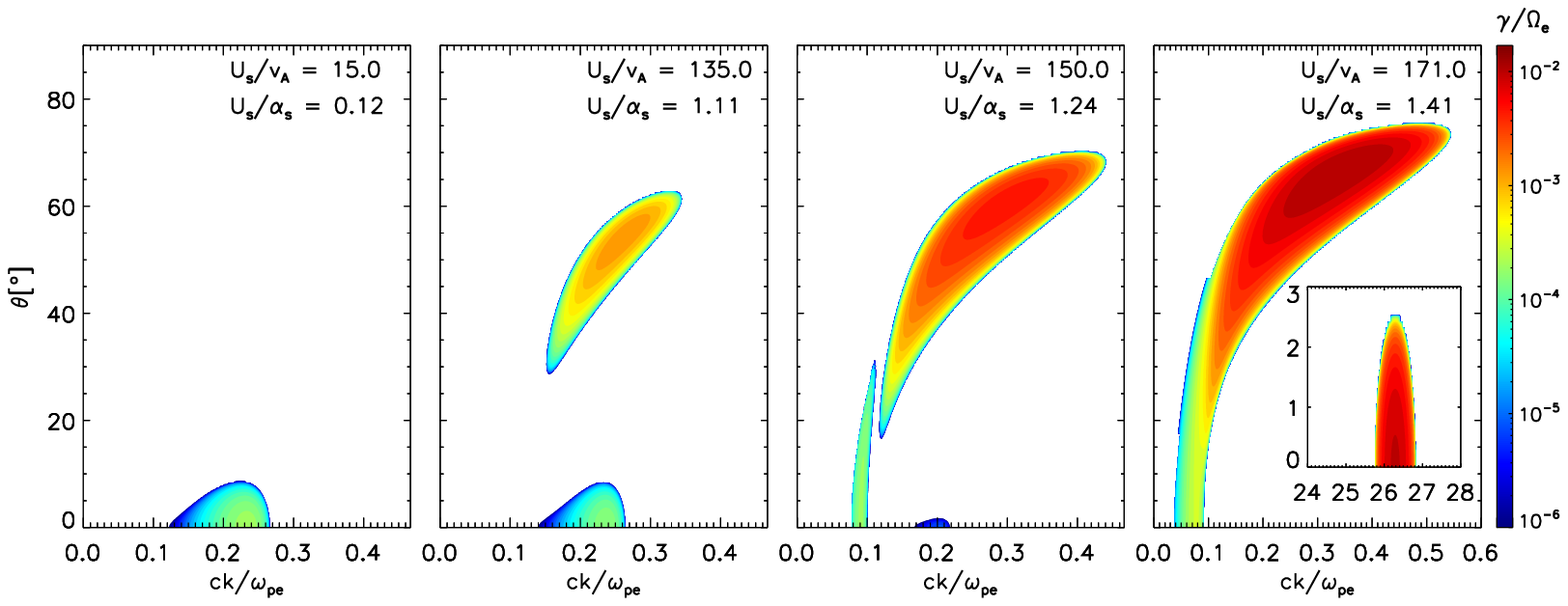}
    \includegraphics[width=0.95\textwidth]{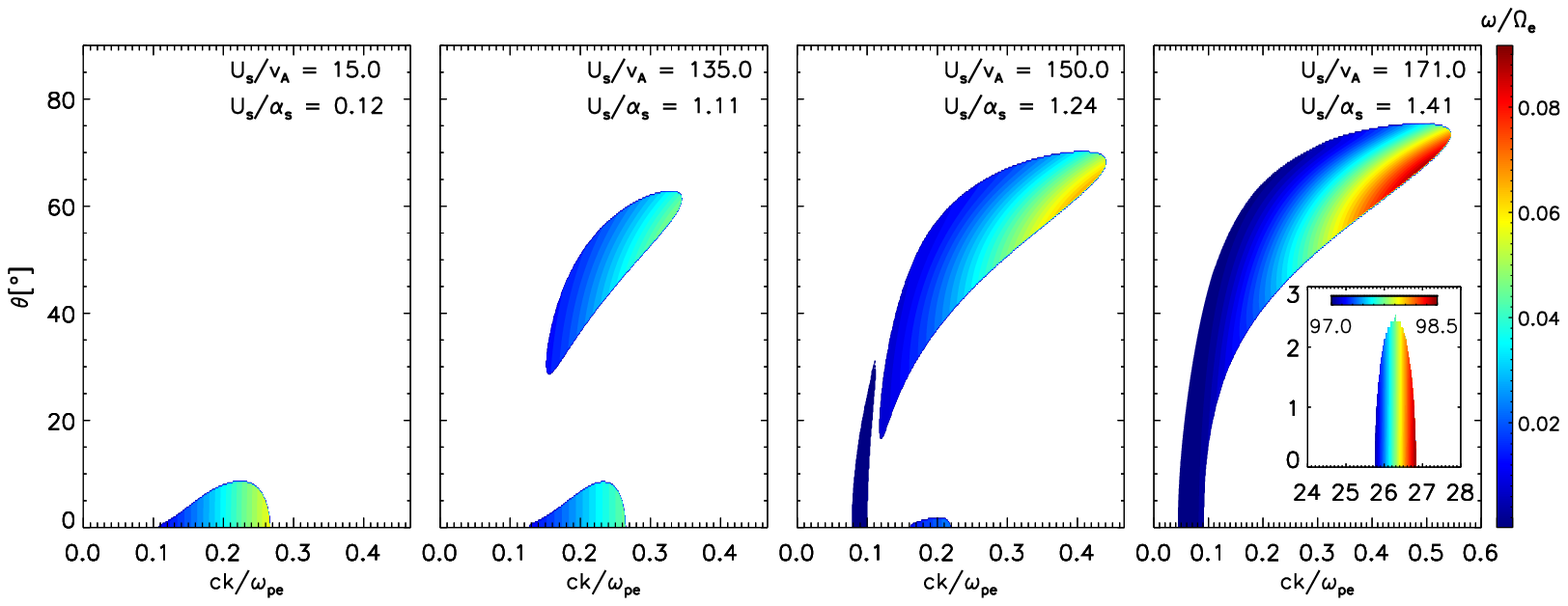}
    \caption{\label{fig1}Growth rates $\gamma/\Omega_e$ (top), and wave-frequency $\omega/\Omega_e$ (bottom), for $\beta_c=2.0$ and various drift velocities, $U_s/v_A=15$,
      $135$, $150$, and $180$.}
  \end{center}
\end{figure*} 

We consider a collisionless quasi-neutral plasma of protons and two electron 
populations, namely, a dense central or core component (subscript ``$c$'') and 
a tenuous suprathermal population (subscript ``$s$'') counter-drifting along the 
ambient magnetic field, assumed constant over at least a few maximum wave-lengths 
of the instabilities considered here (e.g., \citealt{Shaaban-Lazar-2020} and 
references therein)
\begin{equation}
  f_e\left({ v_{\perp},v_{\parallel},}\right) =
  \frac{n_{c}}{n_e}~f_{c}\left({ v_{\perp}, v_{\parallel}}\right)
  +\frac{n_{s}}{n_e}~f_{s}\left({v_{\perp}, v_{\parallel}}\right)\,,
\end{equation}
where $n_e\approx n_p$ is the total electron number density, and $n_{c}$ and 
$n_{s}$ are the number densities of the core and strahl populations, respectively, 
satifying $n_c+n_s = n_e$. In the next this suprathermal 
population will be called generically 'strahl', but the analysis may also apply to 
a core-halo configuration, as explained already above. For both the core ($j=c$) 
and strahl ($j=s$) populations we adopt a simple standard description (widely used 
in similar studies) as drifting bi-Maxwellians \citep{Saito2007b, Verscharen2019}
\begin{equation}
f_j(v_\perp,v_\parallel)=\frac{\pi^{-3/2}}{\alpha_{\perp j}^2\alpha_{\parallel j}}
\exp\left\{-\frac{v_\perp^2}{\alpha_{\perp j}^2}
-\frac{(v_\parallel-U_j)^2}{\alpha_{\parallel j}^2}\right\}\,,
\end{equation}
where $\alpha_{\perp, \parallel,  j}=(2k_BT_{\perp, \parallel, j}/m_e)^{1/2}$ are components of thermal velocities perpendicular ($\perp$) and parallel ($\parallel$) to the background magnetic field, and $U_j$ are drift velocities,  which preserve a zero net current $n_sU_s+n_cU_c=0$. For simplicity, protons are assumed isotropic ($T_{p\perp}=T_{p\parallel}$), nondrifting ($U_p = 0$), and Maxwellian distributed. 

We preset a general dispersion and stability analysis covering the full wave-vector spectrum of (unstable) plasma modes propagating at arbitrary angles $\theta$ with respect to the background magnetic field ($\bm{B}_0=B_0\hat{z}$). Without loss of generality the wave-vector ${\bf k} = k_\perp\hat{x} + k_\parallel\hat{z}$ is chosen in the $x$--$z$ plane ($k_\parallel = k \cos \theta$ and $k_\perp = k \sin \theta$). Our analysis is based on the kinetic Vlasov-Maxwell dispersion formalism, as provided by \cite{Stix1992}, and the unstable solutions are found numerically, providing accurate description for the full spectrum of instabilities (e.g., electrostatic, electromagnetic or hybrid), and various regimes of their co-existence and dominance.
We use a complex root finder based on the M\"uller's method to locate the solutions of the plasma dispersion tensor. Solutions provided by this code have been validated in previous studies for various kinetic instabilities \citep{Shaaban-etal-2019b,Lopez-etal-2019b,Lazar-etal-2019}, and using PIC simulations in the low and high-frequency regimes, and also for multi-component plasmas~\citep{Lopez2017, Lopez2017a, Lopez2020, Micera-etal-2020}.

Present study focuses on the solar wind high plasma beta conditions, i.e., for $\beta_c \gg 0.1$ (more exactly, $\beta_c \gtrsim 1$),  susceptible to various instabilities combining kinetic and reactive free-energy effects of plasma particles. Plasma parameters used in our analysis are tabulated in Table~\ref{t1}, unless otherwise specified. Note that all these values are relevant for the solar wind high-beta conditions, approaching average values reported by the observations, e.g., for the relative number densities of the electron populations, e.g., $n_s/n_e = 1-n_c/n_e = 0.05$, temperature contrast $T_s/T_c = 4$, plasma beta $\beta_c =2$, frequency ratio $\omega_{pe}/|\Omega_e|= 100$ and a realistic proton-electron mass ratio $m_p/m_e = 1836$.
%

\setlength{\tabcolsep}{5pt}
\begin{deluxetable}{lccc}
    \tablenum{1}
    \tablecaption{Plasma parameters used in the present study.\label{t1}}
    \tablehead{& Strahl ($s$)  & Core electrons ($c$) &  Protons ($i$)}
    \startdata
        $n_j/n_i$  & 0.05 & 0.95 & 1.0\\
		$T_{j,\parallel}/T_{i,\parallel}$ & 4.0 & 1.0 & 1.0\\
		$m_j/m_i$ & 1/1836 & 1/1836 & 1.0 \\
		$T_{j, \perp}/T_{j,\parallel}$ & 1.0 & 1.0 & 1.0
    \enddata
    \tablecomments{Other parameters are: $\omega_{pe}/\Omega_e=100$, $\beta_{c}=8\pi n_e T_c/B_0^2=2$}
\end{deluxetable}

We characterize the HFIs as primarily defined by the main plasma eigen-modes destabilized by the relative drift of suprathermal electron population, e.g., (1) fast-magnetosonic/whistler (FM/W) waves, RH-circular  polarized when propagating in parallel direction, (2) Alfv\'enic modes, LH-circular
polarized in parallel direction, and (3) electrostatic beaming instabilities. High beta electrons ($\beta_{\rm eff} = 8\pi n_e k_B T_{\rm eff}/B_0^2= 8\pi k_B (n_c T_c +n_s T_s)/B_0^2 = \beta_c +\beta_s > 0.1$) present in the solar wind are expected to excite moderate and high frequency modes of these branches. The unstable FM/W modes with high frequencies in the range $\Omega_p < \omega_r < |\Omega_e|$ will simply be named whistler heat-flux instabilities (WHFIs), but making however distinction between the (quasi-)parallel and oblique branches of WHFIs \citep{Gary-etal-1994, Wilson-etal-2009, Russel-etal-2009}. 
The instability mechanisms imply resonant or nonresonant interactions with plasma particles, especially electrons, and may determine linear interplay and conversions between different branches of plasma modes. Even in the absence of instabilities, the wave dispersion of electromagnetic (EM) modes decouples from electrostatic (ES) oscillations only for parallel propagation ($\theta = 0$). These aspects will be discussed in the next, in an attempt to accurately identify the regimes of HFIs, and characterize the transition between these regimes. 

\section{Results} \label{sec:results}
\begin{figure}[ht!]
  \begin{center}
    \includegraphics[width=0.48\textwidth]{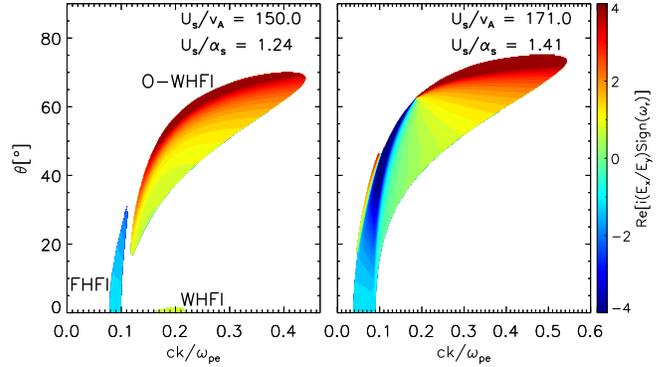}
    \caption{\label{pol}Polarization, $\text{Re}\{i(E_x/E_y)\text{Sign}(\omega_r)\}$, for the last two cases in Fig.~\ref{fig1}, $U_s/\alpha_s=1.24$ (lef) and $U_s/\alpha_s=1.41$ (right).}
  \end{center}
\end{figure} 

We perform a spectral analysis of the unstable modes in ($ck/\omega_{pe}$, $\theta$)$-$space, where $ck/\omega_{pe}$ is the wave-number normalized to the electron inertial length, and $\theta$ is the propagation angle. Upper panels in Figure~\ref{fig1} display the full range of the growth rates $\gamma/|\Omega_e| > 0$ (color codded) derived for different drift velocities of the strahl electrons $U_s/v_A=15$ (left panel), $U_s/v_A=135$ (middle left), $U_s/v_A=150$ (middle right), $U_s/v_A=171$ (right). For a nominal value $v_A = 20$~km/s for the Alfv\'en speed (usually between 10 and 50~km/s at 1~AU) the highest drifts assumed in Figure~\ref{fig1} correspond to the limit values measured for the relative drift of the electron beam/strahl, see  \cite{Wilson-etal-2019b}.
The corresponding wave frequency $\omega/|\Omega_e| > 0$ (color codded) is shown in the lower panels of Figure~\ref{fig1}. The Alfv\'en speed depends only on the ion density and magnetic field, and provides therefore a more neutral normalization, common in the literature. However, here we will also explicitly compare the drift of electron strahl $U_s$ with, $\alpha_s$, the thermal speed of strahl electrons, which is particularly important in the study of kinetic instabilities, directly conditioning their thresholds and dominance regimes, e.g., for the WHFI \citep{Gary-em-1985, Shaaban2018MN} and the electrostatic instabilities \citep{Gary1993}.

\subsection{Whistler heat flux instabilities}

The left panels in Figure~\ref{fig1} describe the (quasi-)parallel WHFI \citep{Gary-em-1985,Shaaban2018MN,Shaaban2018PoP,Tong2019a}, which is solely predicted for the parameters chosen in this case, i.e., less energetic strahls with a low drift $U_s=15 v_A =0.12 \alpha_s < \alpha_s$, lower than thermal speed of the suprathermal drifting electrons. Although the WHFI also extends to small oblique angles, the fastest growing mode propagates in direction parallel to the background magnetic field, i.e., $\theta=0^{\circ}$. These modes are RH circularly polarized, as showin by the positive polarization (green) in Figure~\ref{pol}. Here the polarization is defined as $\text{Pol} = \text{Re}\{i(E_x/E_y)\text{Sign}(\omega_r)\}$, see \citet{Gary1993}. 

\begin{figure}[t]
  \begin{center}
    \includegraphics[width=0.48\textwidth]{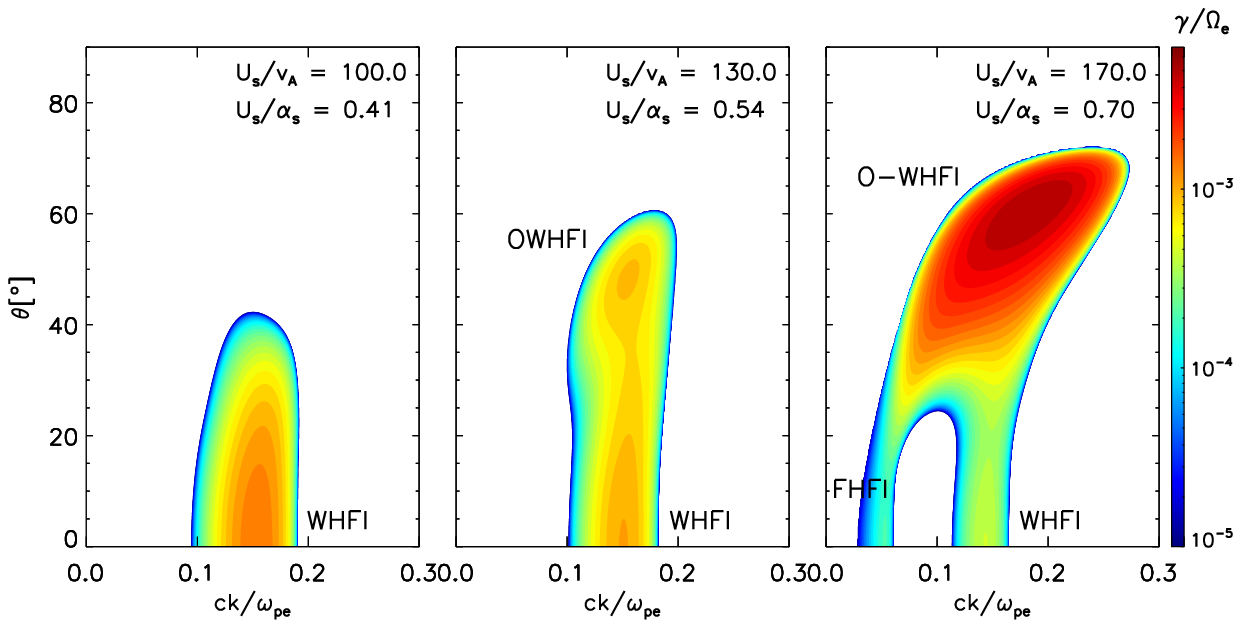}
    \includegraphics[width=0.48\textwidth]{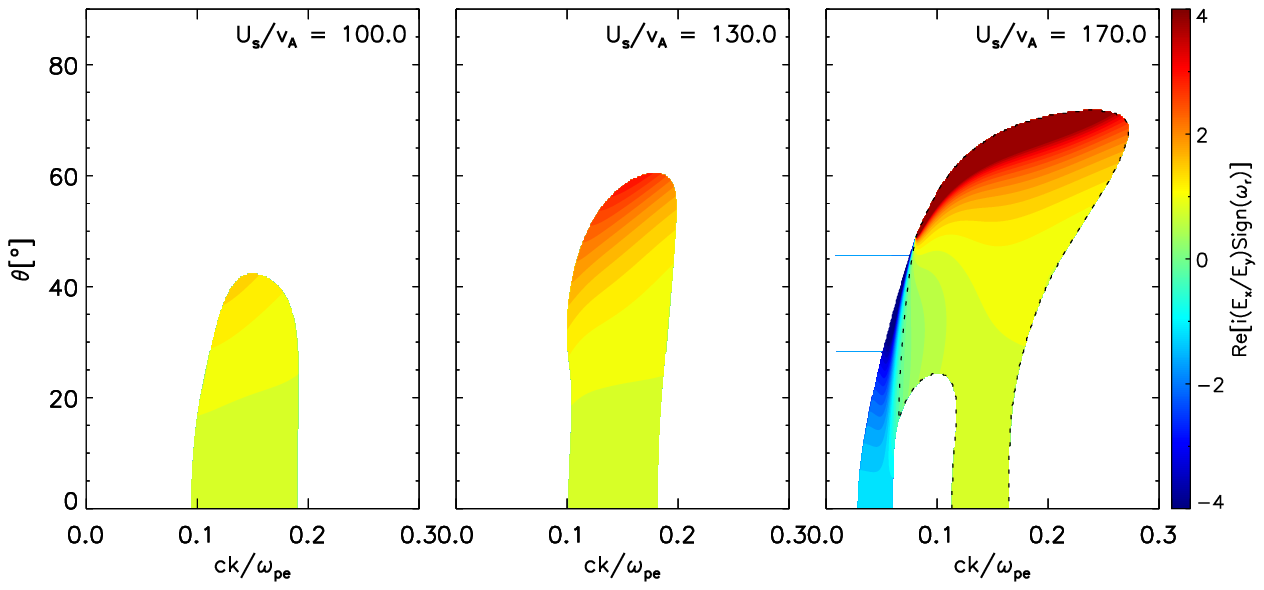}
    \caption{\label{interplay}Transition from the WHFI regime to the dominance of O-WHFI. Growth rate (top) and polarization $\text{Re}\{i(E_x/E_y)\text{Sign}(\omega_r)\}$ (bottom) as a function of wave-number for $\beta_c=8$ and various drift velocities. Dotted black line indicates the contour of minimum polarization ($\simeq 0.0$).}
  \end{center}
\end{figure}

With increasing the drift velocity the growth rate of the parallel WHFI decreases and this mode becomes eventually damped, see  Figure~\ref{fig1}, the middle and right panels, for respectively, $U_s/\alpha_S=$ 1.11, 1.24, and 1.41. Middle-left panels in Figure~\ref{fig1} present the unstable solutions for a higher beaming speed, $U_s=135 v_A = 1.11 \alpha_s$, exceeding the thermal speed. The WHFI restrains, but for oblique angles of propagation we find another whistler-like instability, known already as the oblique WHFI (O-WHFI) \citep{Sentman-etal-1983,  Tokar-etal-1984, Wong-Smith-1994, Verscharen2019}. This oblique mode has a wave frequency dispersion (bottom panels) quite similar to that of parallel whistlers, specific wave-frequencies ($\Omega_p < \omega < |\Omega_e|$) and wave-numbers, and a RH elliptic (positive) polarization for all directions. Polarization is computed (only for the unstable modes, $\gamma>0$) as $\text{Re}\{i(E_x/E_y) \text{Sign}(\omega_r)\}$ and is mapped in Figure~\ref{pol} and bottom panels of Figure \ref{interplay}.
By contrast to the WHFI, the O-WHFI is purely oblique and may reach much higher growth rates. In this case maximum growth rates of the O-WHFI ($\gamma_\text{max}/\Omega_{e}=1.8\times10^{-3}$) are obtained for  $\theta=54.1^\circ$ and $ck/\omega_{pe}=0.26$. The growth rates of this instability are markedly enhanced by only slightly increasing the drift, see the next two cases in Figure~\ref{fig1}. The peaking maximum of the growth rates moves toward higher wave-numbers and larger angles of propagation as the drift velocity increases, i.e., $\gamma_\text{max}/ \Omega_{e}=6.9\times10^{-3}$ at $\theta=60.7^\circ$ and $ck/\omega_{pe}=0.3$ for $U_s/\alpha_s=1.24$, and $\gamma_\text{max}/\Omega_{e}=1.7\times10^{-2}$ at $\theta=66.4^\circ$ and $ck/\omega_{pe}=0.34$ for $U_s/\alpha_s=1.41$.

Figure~\ref{interplay} provides more detail on the gradual transition from the regime of WHFI, predicted in quasi-parallel directions, to the regime dominated by the O-WHFI. The limb of O-WHFI extending to highly oblique angles forms and detaches from the standard WHFI which remains at lower angles. These oblique whistlers can be destabilized by the asymmetric counter-drifting populations of electrons specific to the upstream conditions of the interplanetary shocks \citep{Sentman-etal-1983, Tokar-etal-1984, Wong-Smith-1994}  and to the fast winds \citep{Verscharen2019}. In simulations of a predefined low-scale whistler turbulence the oblique whistlers were found able to strongly interact with strahl electrons, contributing to their pitch-angle and energy scattering \citep{Saito-Gary-2008}. Typical fluctuations of oblique whistlers were also reported by the observations in the magnetosphere during magnetically active periods \citep{Wilson-etal-2011}, in association with electron beams in interplanetary high-$\beta$ shocks \citep{Breneman-etal-2010, Wilson-etal-2012, Ramirez-etal-2012} and recently, collocated with magnetic field holes in the outer-corona \citep{Agapitov-etal-2020}.

\begin{figure}[t!]
  \begin{center}
    \includegraphics[width=0.48\textwidth]{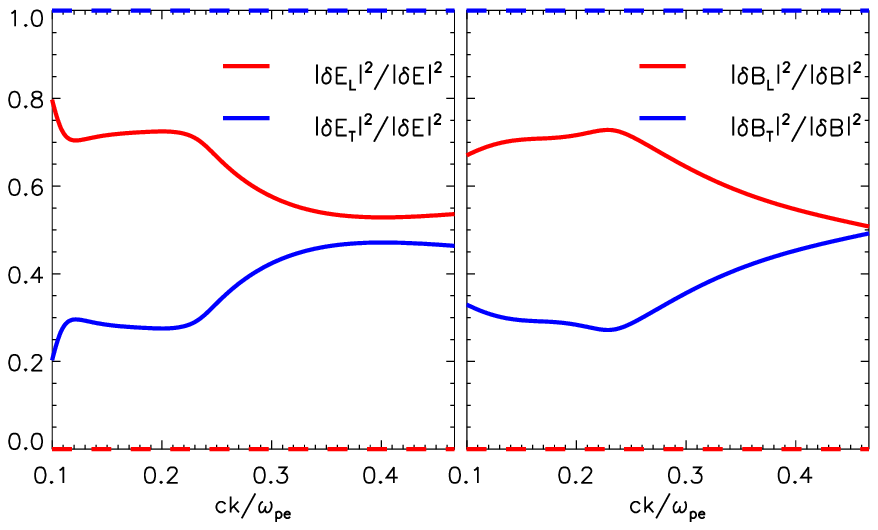}
    \includegraphics[width=0.48\textwidth]{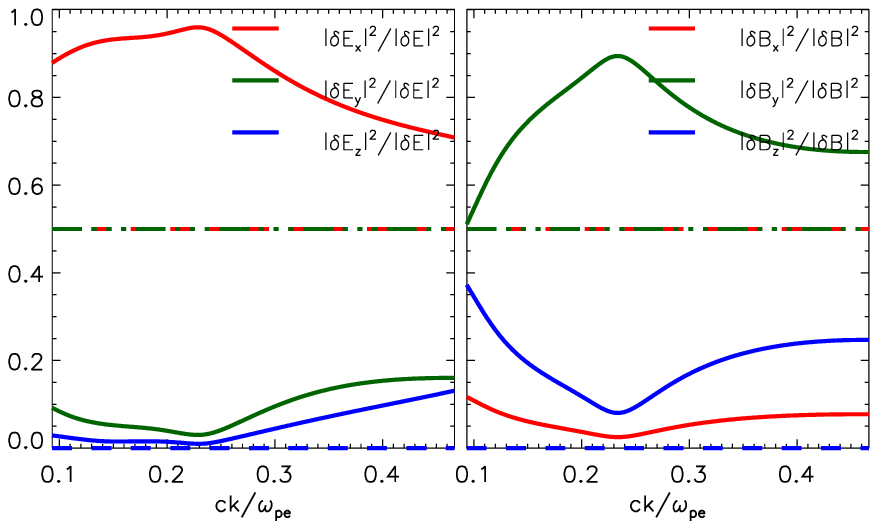}
    \caption{\label{fields}
    Electric and magnetic field powers for the fastest growing O-WHFI, $\theta=60.7^\circ$, in Fig.~\ref{fig1}, third case for $U_s/\alpha_s=1.24$ ($U_s/v_A=150$). Here the directions longitudinal (L) and transverse (T) are defined with respect to the wave-vector, $\delta \mathbf{E}_{L}=(\delta\mathbf{E}\cdot\mathbf{k})\mathbf{k}$. Dashed lines show the magnetic/electric powers of WHFI at $\theta=0^\circ$.
    }
  \end{center}
\end{figure} 

\begin{figure}[t!]
  \begin{center}
    \includegraphics[width=0.48\textwidth]{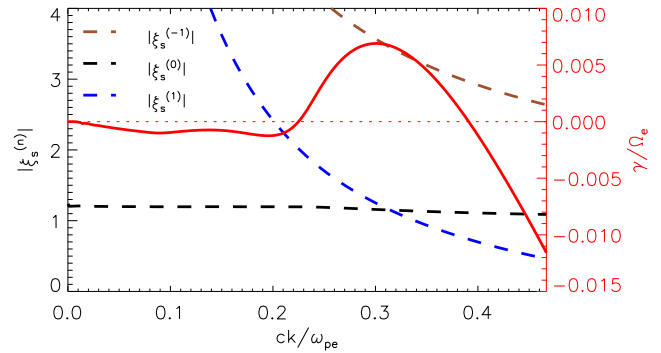}
    \caption{\label{reson} Arguments of plasma dispersion functions $|\xi_s^{(m)}|$ (absolute value) quantifying Landau and transit time resonances $|\xi_s^{(0)}| \to 1$, and cyclotron resonances $|\xi_s^{(\pm1)}| \to 1$, for  the fastest growing O-WHFI in  Figure~\ref{fig1}, third panel. The growth rate is overplotted with a solid red line.}
  \end{center}
\end{figure}

Figure~\ref{fields} displays the wave-number dispersion of the electric and magnetic powers for the fastest growing O-WHFI ($\theta=60.7^\circ$) in Figure~\ref{fig1}, the third case ($U_s/\alpha_s=1.24$). We show the field components in the cartesian $(x,y,z)$ representation (bottom), and with respect to the wave-vector ${\bf k}$, the longitudinal (subscript $L$) or transverse (subscrit $T$) components (top). Dashed lines correspond to the WHFI at $\theta=0^\circ$, as expected for the purely transverse (electric and magnetic) fields propagating in parallel direction.
Based on this understanding, we can claim that O-WHFI can be driven cumulatively by the resonant interactions with beaming electrons, via their Landau and transit time resonances with longitudinal (electrostatic) component $E_L$, and an anomalous cyclotron resonance with transverse (electromagnetic) component $E_T$. The wave-particle resonant mechanisms governing this instability \citep{Tokar-etal-1984} can be identified following the same wave-number dispersion of the arguments of plasma dispersion function (absolute values) $|\xi_s^{(m)}|$, known as ``resonant factors" \citep{Gary-jgr-1975}. These arguments are computed in Figure~\ref{reson} for the fastest growing O-WHFI, the same third case in Figure~\ref{fig1} ($U_s = 1.24 \alpha_s$). The growth rate is overplotted with a solid red line. For wave-numbers corresponding to the maximum growth rate both resonance conditions are well satisfied, i.e., $|\xi_s^{(0)}| \to 1$ involving the Landau and/or transit time resonances, and $|\xi_s^{(\pm1)}| \to 1$ for the anomalous cyclotron resonance. We know already that the anomalous cyclotron resonance can be responsible for the excitation of WHFI, forward propagating modes being overtaken by the strahl electrons \citep{Tokar-etal-1984, Shaaban2018MN}. It is also expected to dominate the mechanism driving O-WHFI at low angles of propagation (mainly involving $E_x$ field component in Figure~\ref{fields}). Instead, highly oblique whistlers are mainly destabilized by the interaction of beaming electrons with the electrostatic and compressive components, through, respectively, a Landau resonance with $E_z$ (which is minor but increases with increasing the wave-number in Figure~\ref{fields}, bottom panels), and a transit time resonance with $B_z$ (which is not minor and shows the similar enhancement with increasing the wave-number in Figure~\ref{fields}, bottom panels). For more explanations see \cite{Gary-jgr-1975}, or the textbook of \cite{Gary1993} and more references therein. 

\subsection{Firehose-like instabilities of Alfv\'enic waves}

Another unstable solution obtained for higher drifts, e.g., the last two cases in Figure~\ref{fig1}, for $U_s/\alpha_s = 1.24$ and 1.41, is the so-called firehose heat flux instability (FHFI). This mode belongs to the Alfv\'enic branch, and in parallel direction it exhibits a maximum growth rate and LH-circular polarization, see also Figures \ref{pol} and \ref{interplay} \citep{Shaaban2018MN, Shaaban2018PoP}. 
The last two cases in Figure~\ref{fig1} show the growth rate (top) and wave-frequency (bottom) of the FHFI, located in a narrow interval of low wave-numbers and low  frequencies. Growth rates are in general lower than those of the O-WHFI, and maximums peak at $\theta = 0^{\rm o}$. New detailed descriptions of the parallel FHFI, including comparisons with the WHFI and the effects of suprathermal electrons present in the solar wind, can be found in \cite{Shaaban2018MN, Shaaban2018PoP}. Last case in Figure~\ref{fig1} ($U_s/ \alpha_s =1.41$) shows the growth rates of FHFI extending to more oblique angles and overlaping with the O-WHFI. However, distinction can easily be made between the LH-polarization of FHFI, i.e., negative values, and the RH-polarization of the O-WHFI, positive values, in Figure~\ref{pol} and \ref{interplay}. Moreover, the O-WHFI is by far dominant, exhibiting much higher growth rates than FHFI. Middle panels in  Figure~\ref{fig1} identify with the regime of dominance of the O-WHFI, when this instability exhibit growth rates much higher than all the other modes, e.g., WHFI or FHFI. However, for higher drifts, e.g., the last case in Figure~\ref{fig1} (for $U_s = 1.41 \alpha_s$), the O-WHFI is already competed by the electrostatic instabilities, showing maximum growth rates for parallel propagation.

\begin{figure}[t!]
  \begin{center}
  \includegraphics[width=0.48\textwidth]{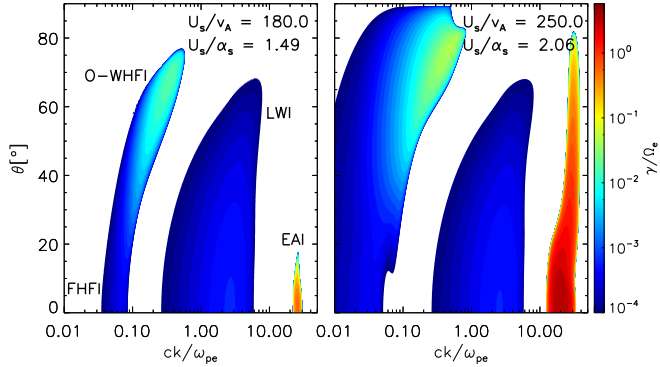}
  \caption{\label{tsi}Linear growth rates, $\gamma/\Omega_e$, $\omega/\Omega_e$, for
      $\beta_c=2.0$, and drift velocities $U_s/\alpha_s=1.49$ (left) and $U_s/\alpha_s=2.06$ (right). 
      }
  \end{center}
\end{figure} 
\begin{figure}[t!]
  \begin{center}
    \includegraphics[width=0.49\textwidth]{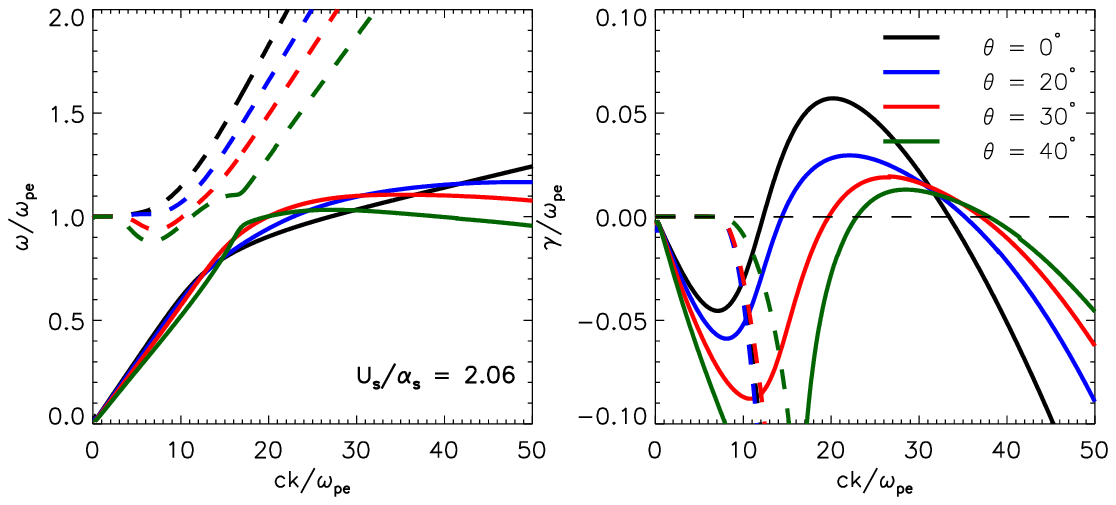}\\
    \includegraphics[width=0.49\textwidth]{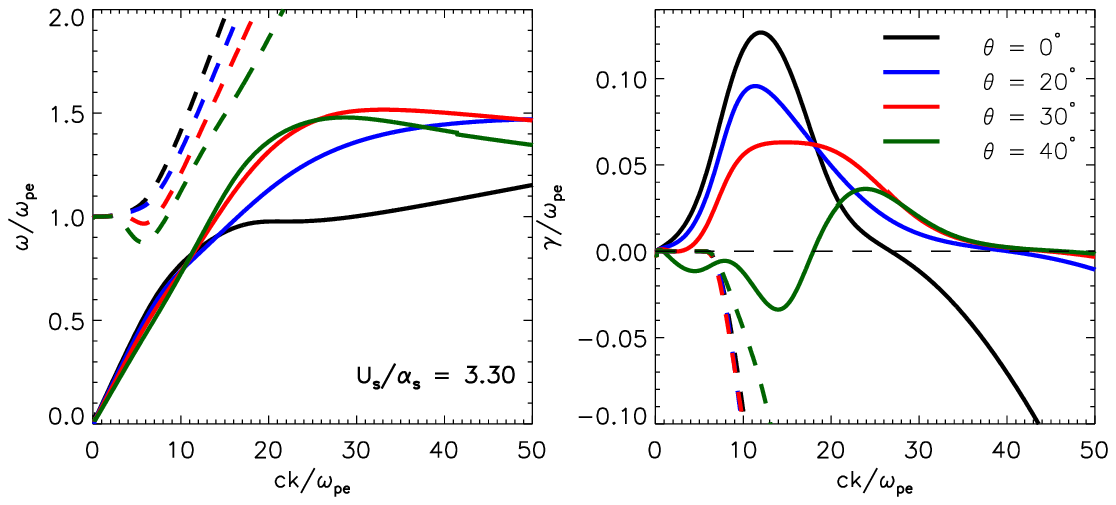}
    \caption{\label{eai} Wave-frequency and growth rate dispersion of the ES instabilities: EAI (solid lines with $\gamma >0$), and LWI (dashed lines with $\gamma >0$), for $U_s/\alpha_s=2.06$ (top), and 3.30 (bottom).}
  \end{center}
\end{figure} 

\subsection{Electrostatic instabilities}

The electrostatic (ES) plasma modes are destabilized when the relative drift of electron strahl is large enough, e.g., $U_s > \alpha_s > \alpha_c$, to ensure Landau resonance with electrons satisfying $\gamma \propto \partial f_s / \partial v_\parallel > 0$.  Thus, the theory predicts a bump-on-tail instability of Langmuir waves for $U_s/\alpha_s <  (n_e/n_s)^{1/3}$, or a more reactive electron beam instability (EBI) for $U_s/\alpha_s > (n_e/n_s)^{1/3}$ \citep{Gary1993}. For highly contrasting electron populations with $T_s > T_c$ the electron acoustic waves become a normal mode, and can be destabilized by a relative core-strahl drift several times higher than thermal speed of the core electrons \citep{Gary1987, Gary1993}.
These instabilities are widely invoked in space plasma applications, to explain electron acoustic emissions detected in the Earth's bow shock \citep{Lin-etal-1985}, radio bursts associated with bump-on-tail instability of coronal or interplanetary shock-reflected electrons \citep{Nindos-etal-2008}, and broadening of solar wind strahls by self-generated Langmuir waves \citep{Pavan-etal-2013} or fast-growing electron beam modes \citep{An-etal-2017, Lee2019}.

\begin{figure*}[t!]
  \begin{center}
    \includegraphics[width=0.95\textwidth]{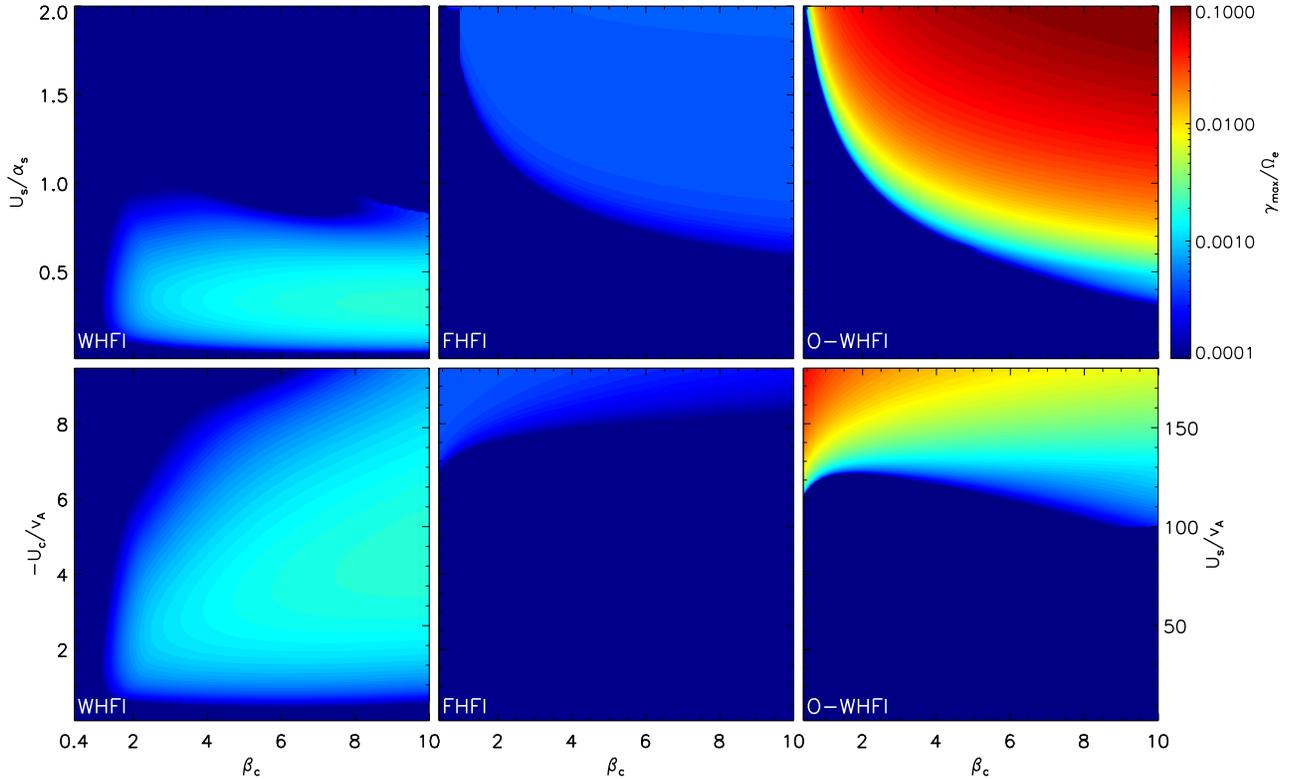}
    \caption{\label{parallel} Maximum growth rates (color codded)  as a function of $\beta_c$ and $U_s/\alpha_s$ (top panels), or $-U_c/v_A$ and $U_s/v_A$ (lower panels) for the WHFI (left), FHFI (middle) and O-WHFI (right panels).}
  \end{center}
\end{figure*} 

The last case in Figure \ref{fig1} shows the electron acoustic instability (EAI) within built-in panels, with growth rates peaking at $\theta = 0^{\rm o}$ ($\gamma_\text{max}/ \Omega_{e}=1.4 \times10^{-2}$) and competing with those of the O-WHFI. In this case drift velocity is $U_s/ \alpha = 1.41 < (n_e/n_s)^{1/3} \simeq 2.71$ and satisfies also conditions for a Langmuir wave instability (LWI - not shown in Figure \ref{fig1}) with growth rates much lower than EAI, see Figure \ref{tsi}. The first panel in Figure \ref{tsi} shows the unstable solutions for a slightly higher drift $U_s/\alpha_s =$1.49, with the EAI in a narrow wavenumber interval but with growth rates much higher than both the O-WHFI and LWI. Note also that FHFI extends to even larger angles but maximum growth rates remain much less than those of the O-WHFI. The LWI and EAI excite waves with frequencies close to the electron plasma frequency ($\omega \sim \omega_{pe} \simeq \omega_{pc}$), but wave-numbers specific to EAI are one order of magnitude higher, see Figures \ref{tsi} and \ref{eai}. 

Figure~\ref{eai} describes the unstable ES modes for $U_s/\alpha_s =$2.06 (top, the same with the right panel of Figure \ref{tsi}), and for  $U_s/\alpha_s =$3.03 (bottom). Specific to more energetic flows and coronal ejections these high drifts are assumed to be at the origin of coronal and interplanetary bursts. In Figure~\ref{eai} we show the wave frequency (left) and imaginary frequency (right) for various angles of propagation, this time normalized by the electron plasma frequency. It becomes thus clear that the fastest growing mode is obtained for parallel propagation, and characteristic frequencies are around the electron plasma frequency. These details enable us to clarify the differences shown by the peaking growth rates in Figure \ref{tsi}. With increasing the drift, maximum growth rates remain in parallel direction, but extend to lower wave-numbers and lower frequencies characteristic to the EBI ($\omega_r \simeq k U_s$). The most unstable modes result from the interplay of EAI and EBI at low angles, and EAI remains solely responsible for the lower growth rates obtained at oblique angles only. In the second case ($U_s / \alpha_s = 3.30$) in Figure \ref{eai} we can distinguish two peaks of the imaginary frequency $\gamma$, which correspond to the EBI and EAI when $\gamma > 0$.  

The opinions regarding the implication of ES, beaming-like instabilities in the regulation of electron strahl are in general divided \citep{Gary-jgr-1975, Pavan-etal-2013, Verscharen2019b}. In this section we clearly show that energetic strahls may provide favorable conditions for these instabilities to develop, identifying the following representative regimes. The last case in Figure \ref{fig1} describes a transition between the O-WHFI and the EAI, when both these two types of fluctuations are expected to interplay. For higher drifts, i.e., in Figure~\ref{tsi}, first panel, the HFIs are dominated by the EAI, while for even higher drifts the second panel in Figure~\ref{tsi} shows another transitory regime from EAI to EBI. The EBI is expected to dominate the unstable regimes for the highest drifts considered in Figure \ref{eai} (bottom panels) satisfying $U_s/\alpha_s > (n_e/n_s)^{1/3} \simeq 2.71$.


\subsection{Drift and beta instability thresholds} \label{sec:overview}

\begin{figure}[t]
  \begin{center}
    \includegraphics[width=0.48\textwidth]{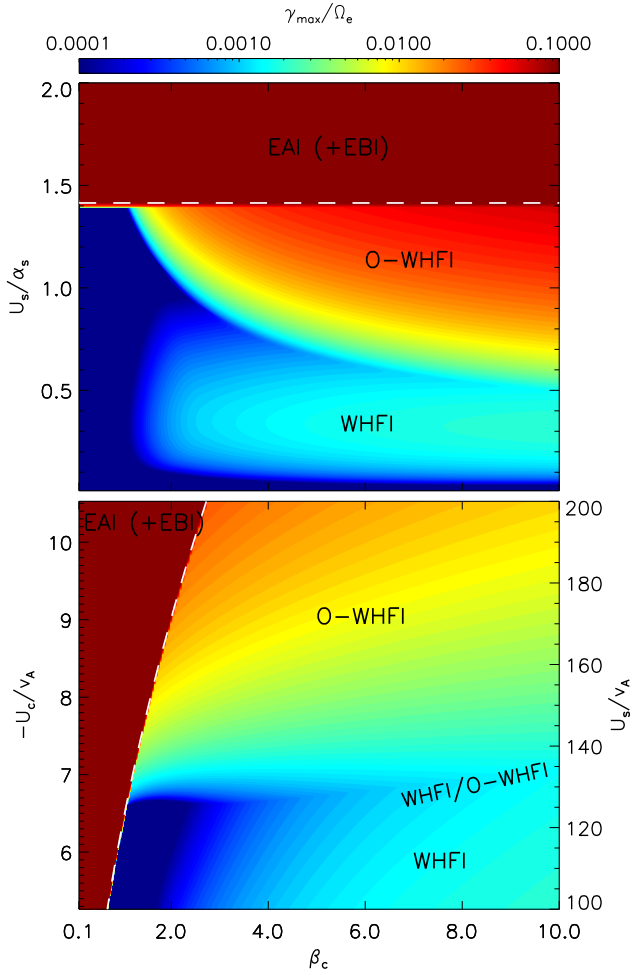}
    \caption{\label{oblique}Maximum growth rate as a function of core plasma beta and beam velocity, $\beta_c$ vs. $U_s/\alpha_s$ (top panel), and $-U_c/v_A$ and also $U_s/v_A$ (lower panel), for all the instabilities discussed, WHFI, FHFI, O-WHFI and EAI (plus EBI). Dashed white line indicates $U_s/\alpha_s=\sqrt{2}$.}
  \end{center}
\end{figure}

We have already identified and characterized a series of alternative regimes of HFIs, as predicted by the theory for different relative drifts of the electron strahl (satisfying the zero net-current condition). The parametric analysis is completed here with a description of the instability thresholds, which highly depend on the electron plasma beta (limiting to high beta conditions, $\beta > 0.1$). Such a general perspective is provided in Figures~\ref{parallel} and \ref{oblique} by the contours of  maximum growth rates $\gamma_\text{max}/\Omega_{e}$, which are  derived in terms of drift velocities for the strahl ($U_s$) or core ($U_c$) and the core plasma beta ($\beta_c$). Note that these contours have no information about $\theta$ or $k$, as they represent the maximum growth rates from the full spectrum of unstable modes (including all frequencies, wave-numbers and angles of propagation) obtained for each combination of drift and electron plasma beta. 

Figure~\ref{parallel} presents contours of maximum growth rates for the WHFI (left), FHFI (middle) and O-WHFI (right panels). These are derived in terms of the core electron beta ($\beta_c$) and the drift velocity, expressed as $U_s/\alpha_s$ (top), or $-U_c/v_A$ and $U_s/v_A$ (lower panels).
There are unstable regimes which appear in both cases, but complementary regimes are also shown, for instance, those hidden by a direct dependence of $\beta_c$ on $v_A$ (via the density and magnetic field) are shown in top panels, while those hidden by a more subtle dependence of $\beta_c$ on $\alpha_s$ (due to a fixed core-strahl temperature contrast $T_c/T_s = 1/4$, see Table~\ref{t1}, leading to $\alpha_c = (T_c/T_s)^{1/2} \alpha_s$) appear in bottom panels. On the other hand, the variations of relative drifts with respect to thermal speed $\alpha_s$ (top panels) may have an extended physical relevance, helping us not only to delimit complementary regimes corresponding to different instabilities, e.g., WHFI from FHFI, or even from ES instabilities, but to understand the difference between physical mechanisms responsible for these instabilities (as discussed already above). 

Left panels in Figure~\ref{parallel} show a non-monotonous variation of the growth rate of WHFI with the drift velocity, as the growth rate increases and then decreases with increasing the drift. Consequently, the most unstable solutions of WHFI are located in-between the lower and upper thresholds, as also found by  \cite{Shaaban2018MN, Shaaban2018PoP} for lower $\beta_c \lesssim 1$ plasma conditions. 
%
Complementary to WHFI, for higher drifts the theory predicts two distinct instabilities. Middle panels of Figure~\ref{parallel} show  the maximum growth rates of FHFI, with a monotonous variation with the drift velocity, and the core plasma beta ($\beta_c$). The maximum growth rate $\gamma_\text{max}/ \Omega_{e}$ of FHFI increases with the drift velocity, but decreases as $\beta_c$ increases (bottom panels). The most unstable FHFI is located at large drifts ($U_c$) and low $\beta_c$. Secondly, right panels in Figure~\ref{parallel} show the  O-WHFI, mostly overlapping with the parametric regime of FHFI, but the O-WHFI exhibits much higher maximum growth rates than FHFI and WHFI. Similar to FHFI, the maximum growth rate of the O-WHFI is, in general, a monotonous function of the drift velocity and core plasma beta. The O-WHFI is stimulated by increasing the drift velocity and decreasing the core plasma beta. For low beta the most unstable O-WHFI is located at large drifts, but with increasing the plasma beta this instability becomes operative for lower drift velocities. The lowest drifts remain susceptible only to WHFI.

\begin{figure}[t]
  \begin{center}
    \includegraphics[width=0.48\textwidth]{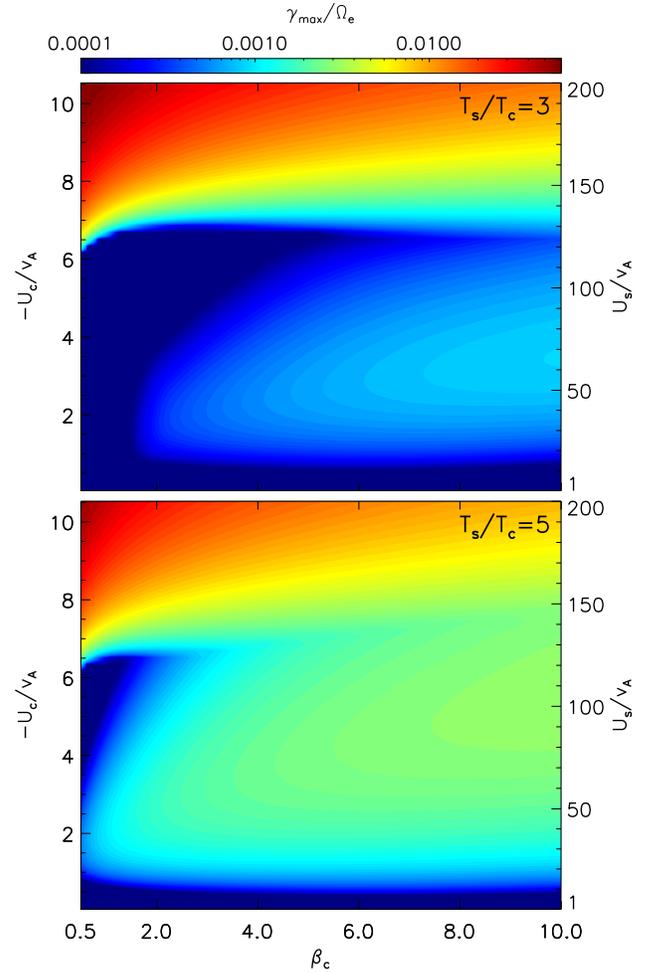}
    \caption{\label{rtemp}Maximum growth rate as a function of core plasma beta and beam velocity, $-U_c/v_A$ and $U_s/v_A$, for different temperature ratios between core and strahl, $T_s/T_c=3$ and $5$.}
  \end{center}
\end{figure} 

The alternative regimes of EM instabilities described in Figure~\ref{parallel} are contrasted in Figure~\ref{oblique} with the very high growth rates of ES instabilities. The range of plasma beta is extended to the interval $0.1\leq\beta_c\leq10$, to include lower beta conditions. 
For moderately high values of beta (e.g., $\beta_c = 2$), WHFI and O-WHFI are complementary, their regimes, respectively, at the lowest or higher drifts velocities, being well delimited by the lowest contour levels of $\gamma_{\rm max}$. For higher values of $\beta_c$ these two regimes overlap, in-between defining a transition where WHFI and O-WHFI interplay and may compete to each other.
The lower beta part of the figure is dominated by the ES instabilities, which involve the EAI and for higher drifts the EBI. These instabilities exhibit very high growth rates, which explains the abrupt transition to the O-WHFI. Marked with white-dashed lines at about $U_s/\alpha_s \simeq \sqrt{2}$, these narrow threshold conditions are characteristic to the interplay of O-WHFI and EAI described in the last case of Figure \ref{fig1}. For our parameterization characteristic to the solar wind, the growth rate of FHFI is always smaller than the O-WHFI or EAI, and we could not find any regime where FHFI can develop.

Finally, in Figure~\ref{rtemp} we show the effect of the strahl-core temperature ratio, contrasting maximum growth rates obtained for lower and higher values, respectively, $T_s/T_c=3$ in the top panel and $T_s/T_c=5$ in the bottom panel. These values are in the range of solar wind measurements \citep{Wilson-etal-2019b}. Major differences are observed for the WHFI thresholds. For higher ratios $T_s/T_c$, the region of dominance of the WHFI extends to lower betas and higher drifts (as also shown in Figure~\ref{parallel} for the case $T_s/T_c=4$) covering a larger portion of the parameter space. The region of co-existence of WHFI and O-WHFI also extends to lower values of $\beta_c$, while the O-WHFI region of dominance remains almost unchanged, although the maximum growth rates of this instability decrease as the temperature ratio increases.

\section{Conclusions} \label{sec:conclusions}

We have provided a unified description of the full spectrum of heat-flux instabilities (HFIs) driven by the relative drift of suprathermal electron populations under the high-beta solar wind conditions. Their nature, wave dispersion, stability and polarization highly depend on the relative drift (or beaming) velocity of suprathermal electrons and the plasma beta parameter. The zeroth order counter-drifting distributions are modeled with standard drifting-Maxwellians, which enable simple parameterizations and straightforward analysis and interpretation of HFIs in various conditions typically encountered in the solar wind, e.g., a drifting halo in the slow wind, or the electron strahl carrying the heat flux of the high speed flows.

The unstable solutions have been derived and examined in terms of their main features, i.e., wave frequencies, growth-rates, wave-numbers and propagation angles, and in terms of plasma (electron) parameters defining the instability conditions, thresholds, etc. Predicted are three electromagnetic instabilities, namely, the quasi-parallel whistler heat-flux instability (WHFI), the firehose heat-flux instability (FHFI) and the oblique WHFI (O-WHFI), and a series of electrostatic instabilities destabilizing Langmuir waves (LW), the electron acoustic (EA) modes, or the more reactive electron beam instability (EBI).

We can identify three alternative regimes, each of them characterized by a well defined instability, solely predicted by the theory or with (maximum) growth rates much higher than other unstable modes. Thus, for relatively low drifts of  suprathermal electrons, i.e., $U_s< \alpha_s$, the WHFI is the only operative, with maximum growth rates associated with parallel propagation. This regime is characteristic to the low drifts and the large quasithermal spread of halo electrons, and seems to be controlled exclusively by the WHFI \citep{Gary-em-1985,Shaaban2018MN, Scime1994}. Typical WHF fluctuations associated with drifting suprathermal populations are confirmed by the solar wind observations, see \cite{Wilson2013, Tong2019a,Tong2019b}.

For higher drifts the dominance shifts to the O-WHFI, which are hybrid modes triggered unstable by cyclotron resonance, mainly at small angles of propagation, combined with Landau and transit time resonances, dominant at larger angles \citep{Sentman-etal-1983,  Tokar-etal-1984, Wong-Smith-1994, Verscharen2019}. With increasing the drift the instabilities become more specific to a core-strahl configuration, switching from a kinetic nature  near the threshold to a more reactive type for higher drifts. The growth rate of O-WHFI increases with the drift, and is in general higher (or even much higher) than that of the WHFI. The wave fluctuations resembling oblique whistlers are indeed associated with electron beams in the solar wind observations \citep{Breneman-etal-2010, Wilson-etal-2012, Ramirez-etal-2012}.
The increase of temperature contrast ($T_s/T_c$) slightly inhibits the growth rates of O-WHFI, but stimulates the WHFI extending the instability conditions to lower plasma betas. The effect is similar to that caused by a decrease of the relative drift, leading to a regime more specific to the halo electrons. 

Theoretically, the electrostatic modes can be destabilized already for $U_s > \alpha_s$, for instance, conditions for a bump-on-tail instability can be satisfied for $\alpha_s <  U_s < (n_e/n_s)^{1/3} \alpha_s (\simeq 2.71 \alpha_s)$ to excite Langmuir waves. However, in the given conditions these modes may not have any chance to develop because their growth rates are much lower than those of the O-WHFI predicted for the same conditions.  Instead, due to the temperature contrast between electron populations another instability is predicted, namely, the EAI triggered by drifting electrons with $U_s \geqslant \sqrt{2}\alpha_s$ (when $\beta_e =2$). Near the threshold this instability strongly compete with the O-WHFI, while for slightly higher drifts the growth rates of EAI become already very large, with peaking values at least one order of magnitude higher than those of the O-WHFI. For even more energetic beams satisfying $U_s > (n_e/n_s)^{1/3} \alpha_s (\simeq 2.71 \alpha_s)$, which are relevant for the fast outflows in the outer corona (also coronal mass ejections), the theory predicts an additional EBI. Near the threshold of this instability we found four unstable modes, O-WHFI, LWI, EAI ad EBI, but only the last two have chances to develop, with comparable growth rates much higher than those of the other two modes. These kind of electrostatic instabilities are widely invoked in space plasma applications, but the resulting high amplitude fluctuations may undergo rapid nonlinear decays and are ultimately witnessed by the electromagnetic or radio emissions, see \cite{Nindos-etal-2008} and refs therein.

Summarizing, our results identify three complementary regimes of HFIs, associated to three distinct instabilities, the parallel WHFI, the O-WHFI or the EAI, and interlinked by a series of transitory regimes. For each transition the theory predicts the interplay or co-existence of at least two distinct instabilities, for instance, the interplay of parallel and oblique whistlers for lower drifts, and with increasing the drift a mixing the O-WHFI and EAI, or the limit case where the EAI and EBI can develop concomitantly. These findings strongly suggest that heat flux regulation may not involve only one but several HFIs, concomitantly or successively in time. Conditions for a single, well defined instability with major effects on the suprathermal electrons and, implicitly, the heat flux, may be very limited. Whistler HFIs are more likely to occur but only for minor drifts (as also reported by recent observations), which may explain a modest implication in their regulation, shown already in quasilinear studies \citep{Shaaban-etal-2019, Shaaban-etal-2019b} and numerical simulations \citep{Lopez-etal-2019}.

We can finally conclude stating that a realistic plasma parameterization combined with a selective spectral analysis can be crucial for understanding the nature and origin of HFIs and their implication in the regulation of the solar wind heat flux. Our theoretical predictions are expected to stimulate further investigations using full kinetic simulations, and confirm the existence of these alternative regimes, not only in the initial linear phase of HFIs but also during their quasi- or non-linear growth in time, which involves a relaxation of the relative drift and, implicitly, changes to different successive regimes of HFIs corresponding to lower drifts.

\acknowledgments
These results were obtained in the framework of the projects SCHL 201/35-1 (DFG-German Research Foundation), C14/19/089  (C1 project Internal Funds KU Leuven), G.0A23.16N (FWO-Vlaanderen), and C~90347 (ESA Prodex). R.A.L thanks the support of AFOSR grant FA9550-19-1-0384. S.M. Shaaban acknowledges support by a FWO Postdoctoral Fellowship, grant No.~12Z6218N. P.S. Moya is grateful for the support of KU Leuven BOF Network Fellowship NF/19/001, and ANID Chile through FONDECyT Grant No. 1191351.


\end{document}